# Foundation Models for Medical Imaging: Status, Challenges, and Directions

Chuang Niu, *Member, IEEE*, Pengwei Wu, *Member, IEEE,* Bruno De Man, *Fellow, IEEE*, and Ge Wang, *Fellow, IEEE*

*Abstract*— **Foundation models (FMs) are rapidly reshaping medical imaging, shifting the field from narrowly trained, task-specific networks toward large, general-purpose models that can be adapted across modalities, anatomies, and clinical tasks. In this review, we synthesize the emerging landscape of medical imaging FMs along three major axes: principles of FM design, applications of FMs, and forward-looking challenges and opportunities. Taken together, this review provides a technically grounded, clinically aware, and future-facing roadmap for developing FMs that are not only powerful and versatile but also trustworthy and ready for responsible translation into clinical practice.**

## I. INTRODUCTION

Artificial intelligence (AI) for medical imaging is experiencing a transformative shift from task-specific models toward foundation models (FMs), which are large artificial neural networks pre-trained on vast, diverse datasets and adapted efficiently to a variety of downstream tasks. In medical imaging, where labels are scarce, heterogeneous, and expensive, FMs show a strong promise for rapid adaptation with minimal annotation, improved generalization across sites, scanners, and populations, and a plausible route to "generalist" medical imaging assistants that reason across contexts.

Recent overviews from both the radiology and computer vision communities document a surge of FM research, spanning 2D/3D segmentation, image–text representation learning through vision–language fusion, and generative models. Together, these developments motivate a new synthesis of principles, capabilities, and translational considerations tailored to the healthcare ecosystem [1].

To contextualize foundation models, we begin by exploring their relationship with the broader AI landscape, coupled with Figure 1.1 illustrating the relative timelines of the related areas along with some seminal publications. AI refers to non-human systems performing tasks that mimic human perception and reasoning, such as language understanding and image analysis. Machine learning, a subset of AI, trains models to detect patterns in data, evolving from simple statistical methods to more sophisticated tools like random forests and support vector machines. Deep learning uses multi-layer artificial neural networks to represent data in a data-driven fashion, leading to advanced architectures, like Convolutional Neural Networks (CNNs), Recurrent Neural Networks (RNNs), Graph Neural Networks (GNNs), and Transformers.

The term *foundation model* was coined by the Center for Research on Foundation Models at the Stanford Institute for Human-Centered Artificial Intelligence in August 2021 [19]. Foundation models are a class of deep learning models that are initially trained based on a diverse dataset for broad applications, and that can then be fine-tuned for specific downstream applications. Typically, they are initially trained in a self-supervised fashion. These pre-trained FMs then serve as the basis for developing task-specific models through transfer learning. The term foundation model is sometimes used loosely: a critical examination of the criteria for a model to qualify as a foundation model is given in [20].

Foundation models are characterized by enormous training data and parameter counts, which lead to emergent capabilities that do not present in smaller models. In other words, a foundation model serves as a general-purpose platform that, with minimal task-specific training, can achieve strong performance across a variety of tasks. Another hallmark of foundation models is scalability. Their performance improves predictably as model size, training data, and amount of compute increase, following empirical scaling laws. This scaling yields surprising capabilities, e.g., GPT-3 demonstrated in-context learning to solve tasks it was not explicitly trained for. Foundation models also exhibit strong generalization and transferability, meaning that the knowledge captured during pretraining on broad data can be transferred to unseen tasks. A single pretrained model can be fine-tuned to excel in applications ranging from natural language processing (NLP) to computer vision and robotics. This versatility has incentivized homogenization of AI research around a few architectures, especially the Transformer. However, this also means any defects or biases in a foundation model might propagate to its downstream uses.

We first introduce several previous review papers related to foundation models. A comprehensive survey of self-supervised learning (SSL) is provided by Liu et al in 2021 [21]. Two 2023 surveys by Kazerooni et al. and by Yang et al [22], [23] provide in-depth overviews of the rapidly evolving field of diffusion models, which are increasingly being integrated into foundation models. A more recent review of generative models is provided by Hein et al [24]. Longpre et al. [25] present a practical guide to support responsible and transparent development of FMs across text, vision, and speech modalities.

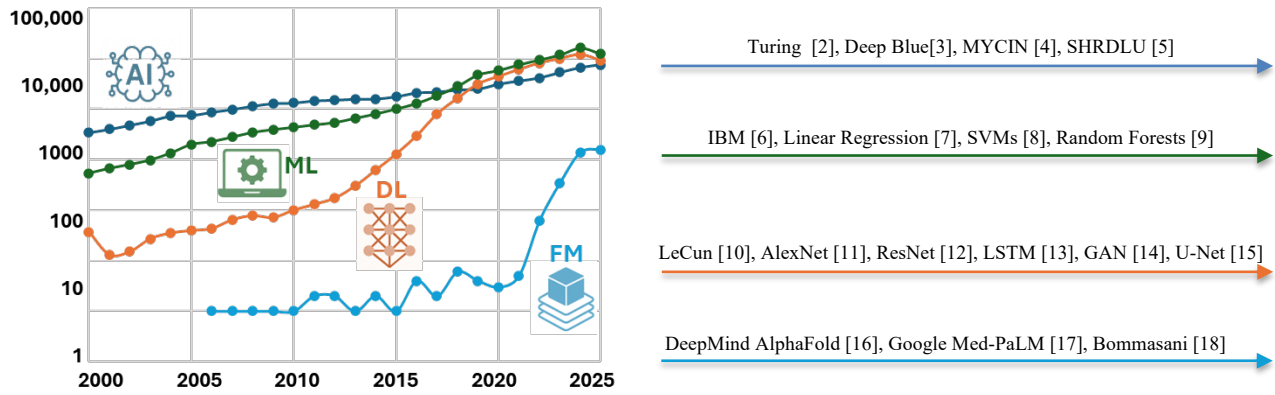

Fig. 1.1: Publication trends from 2000 to 2025 (Source: Scopus, Sep' 2025) highlight the relative growth of Artificial Intelligence (AI), Machine Learning (ML), Deep Learning (DL), and Foundation Models (FM), alongside key milestones that shaped each field.

Large language models (LLMs) are the most popular type of FMs. Zhou et al. [26] trace the evolution from BERT to ChatGPT, emphasizing key advancements in architecture, training methods, and model capabilities. Zhao et al. [27] summarize LLMs and emerging trends like multi-agent collaboration and chain-of-thought reasoning. Ian A. Scott [28] introduces physicians to FMs and LLMs, explaining how they can perform diverse tasks across modalities (text, audio, images, video), with potential applications in medicine. Yang et al. [29] offer an overview of ChatGPT, BERT, and other LLMs, detailing their underlying architectures, training strategies, and broad applications. Truhn et al. [30] explore how LLMs and multimodal foundation models are transforming precision oncology.

Some excellent reviews focused on medical imaging and image analysis. Azad et al. [31] explore how FMs are reshaping the field of medical imaging, including a structured taxonomy of FMs in medical imaging and clinical applications, challenges, and directions. The vision-language models (VLMs) are covered in several reviews, including Huang et al. [32], Ryu et al [33], and Sun et al [34]. They analyze how multimodal FMs are reshaping clinical AI by integrating visual data (e.g., X-rays, MRIs) with textual information (e.g., radiology reports, clinical notes). Zhang et al [35], Huix et al [36], and Veldhuizen et al [37] survey FMs for medical image analysis and outline unique challenges of applying them in radiology, pathology, and ophthalmology. Khan et al [38] present an analysis of subgroup fairness in medical imaging FMs, indicating that improved overall accuracy may come at the expense of reduced subgroup fairness.

The latest milestone in this emerging field is the IEEE Transactions on Medical Imaging Special Issue on Advancements in Foundation Models for Medical Imaging (2025), which assembled 18 papers spanning segmentation, multimodal integration, architectural innovations, benchmarking, ethics, and generative synthesis. Collectively, these contributions underscore both the breadth and depth of current progress: from SAM-inspired segmentation frameworks and Mamba-based backbones to multimodal vision–language adaptations and large-scale echocardiography models; from topology-guided generative pathology models to benchmark and ethical analyses that foreground fairness, interpretability, and governance. The Special Issue illustrates not only rapid technical advances but also the broader community recognition that foundation models for medical imaging and beyond must be judged by accuracy, equity, transparency, and clinical utility. This collection thus provides a valuable snapshot of the state of the art, while also motivating the need for integrative up-to-date reviews like the present article.

This review advances the current literature in three distinct ways. First, it adopts a broad coverage of FMs in medical imaging, especially incorporating the underrepresented domain of image reconstruction for CT, SPECT, PET, MRI, ultrasound, and optical imaging. Second, by integrating the most recent developments in this rapidly evolving field, our review addresses temporal gaps in prior surveys, such as generative AI, reinforcement learning, and modern reasoning methods for medical imaging researchers and practitioners. Finally, we offer an extensive perspective as the last part that reflects our current vision to promote further advancement.

The remainder of this review is organized as follows. In the next section, we distill the principles behind FMs that are most relevant to imaging, ranging from major model architectures, common training strategies, to the key components of development and deployment of FMs. In the third section, we survey applications across medical imaging modalities (CT, MR, PET, US, X-ray, ophthalmology, pathology) and tasks (segmentation, detection, diagnosis, triage, report generation, reconstruction), highlighting strengths, caveats, and challenges. In the final section, we identify future directions in terms of four pillars supporting medical imaging FMs, which are data/knowledge, model/optimization, computing power, and regulatory science. Overall, we hope to provide a unifying view that is technically grounded, clinically actionable, and forward-looking.

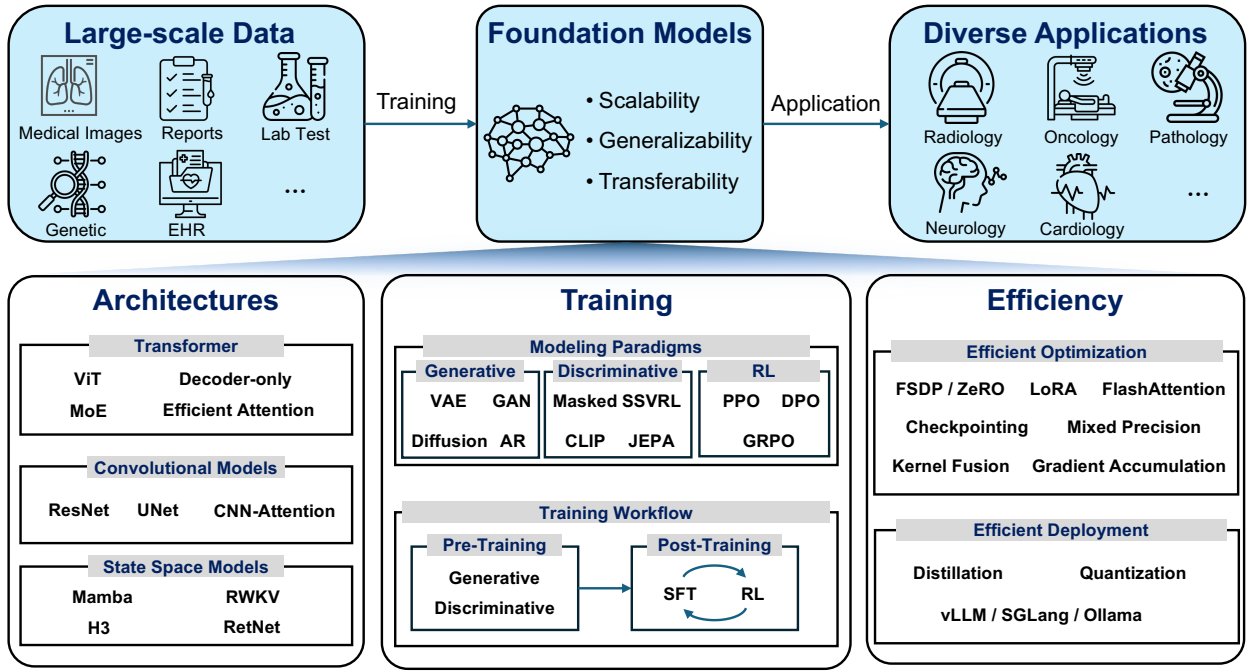

Fig. 2.1. Principles of Foundation Models. Overview of medical foundation models, illustrating how large-scale heterogeneous clinical data, including medical images with associated reports, lab tests, genetics, and electronic health records, are used to train scalable, generalizable, and transferable foundation models that can be adapted to diverse downstream applications such as radiology, oncology, pathology, neurology, and cardiology. The lower panel summarizes the major technical components, including model architectures, modeling paradigms and training workflows, and efficiency techniques for optimization and deployment.

## II. PRINCIPLES OF FOUNDATION MODELS

In medical imaging, FMs learn from large-scale, image-centric multimodal datasets along with associated radiology reports, laboratory results, genetic profiles, and electronic health record (EHR) data. These models support a wide range of clinical specialties, including radiology, oncology, pathology, neurology, cardiology, and so on. This section outlines the core principles behind medical imaging FMs, covering model architectures, training strategies, and efficiency techniques, as summarized in Figure 2.1.

### A. Model Architectures

Several neural network architectures serve as the building blocks for foundation models. Briefly speaking, Transformers have taken the lead in many domains, especially in NLP and image analysis. CNNs, however, often outperform Transformers on smaller-scale vision tasks when data is scarce, due to their built-in locality bias. For some tasks with very long sequential data, state space models (SSMs) like Mamba now show promising results, even surpassing Transformers of similar or larger sizes. In this subsection, we review the major architectures, their variants, highlighting strengths and limitations.

#### 1) Transformer

The Transformer [39] has become the de facto core of most FMs in language and increasingly in vision and medical imaging. Transformers dispense with recurrence and convolutions in favor of self-attention mechanisms that explicitly model long-range dependencies. A Transformer block is typically composed of multi-head self-attention layers and feed-forward layers, enabling it to attend to all positions of an input sequence in parallel.

Vision Transformers (ViTs) [40], [41] tokenize images and analyze them based on sufficient training data. ViTs can achieve excellent performance across various vision tasks. ViTs' strengths lie in their ability to capture global context easily via self-attention and their scalability with gradually refined attention coverage. However, since Transformers lack inductive biases, a ViT trained from scratch on limited data may underperform a CNN [41]. The Swin Transformer [42] was designed to address the ViT's issues by computing self-attention in non-overlapping windows and shifting the window positions between layers to allow cross-window connections to reduce the computation and improve generalization on smaller datasets. Another strategy of reducing the computation for high-resolution/dimension images is to interleave global and local window attention across layers [43].

Decoder-only Transformers [44] are simplified Transformers where only the decoder stack is used. Through causally masked self-attention, each token can attend only to preceding tokens, making them inherently autoregressive. Importantly, causal self-attention is powerful in deployment due to the ability to perform key-value caching, which brings several tangible advantages, such as huge speedup in inference and stable latency. Thus, it has become foundational for modern language models. The decoder only architecture is also critical in multimodal modeling by decoding vision and language tokens for various tasks, such as image/video captioning [45] and medical report generation [46].

The Mixture of Experts (MoE) architecture [47] is an enhancement to Transformer architectures and replaces the standard feed forward layer in each Transformer block with a set of parallel sub networks called experts and a learnable gating network that routes each input token to a small subset of experts based on token-specific features, dramatically increasing model capacity by only activating a fraction of the total parameters per token. Each gate typically employs a sparsely gated softmax function along with auxiliary load balancing losses and bias terms to distribute tokens evenly across experts [48]. This architecture enables training models with trillions of parameters, where only a few billion of those parameters are used during inference for any given token, making MoE an enabler of scalability in modern FMs.

Efficient attention mechanisms address the quadratic cost of standard Transformers by reducing computation and memory while preserving essential context. Sparse attention limits each token's receptive field through structured or adaptive sparsity [49]. Linear attention replaces softmax with kernelized or alternative formulations so attention scales linearly in sequence length [50], [51]. Low-rank and factorized approaches compress keys, values, or the attention matrix itself [52]. Meanwhile, multi-query and group-query attention [53], [54] share key–value projections across heads or head groups, boosting memory efficiency and throughput with minimal loss in expressiveness. Together, these methods form a toolkit that enables scalable Transformers for long-context, high-resolution, and multimodal tasks.

### 2) Convolution-based Models

Convolutional Neural Networks (CNNs) [55] dominated computer vision for years, and they remain highly relevant in the era of FMs. CNNs like ResNet [56] and UNet [15] are good at learning local patterns and are translation-invariant. These inductive biases allow CNNs to generalize well with relatively small training datasets and excel on medium-scale tasks with strong local features. However, CNNs have a restricted receptive field, so localized convolutions might miss global context. Attention-Convolutional Models [57] aim to get the best of both CNNs and Transformers, such as by augmenting CNN backbones with attention blocks [58], or by augmenting Transformers with convolutional token embeddings [59]. However, they could also inherit some limitations of both paradigms. Nonetheless, these hybrid models form an important class of FM architecture.

### 3) State-Space Models

Recurrent Neural Networks (RNNs) [60] were the workhorse for sequence modeling. However, RNN and its variants like Long Short-Term Memory (LSTMs) and Gated Recurrent Unit (GRUs) [61] faced challenges in capturing long-range dependencies. In particular, they could not be parallelized across sequence positions. To this end, State space models (SSMs) provide a powerful framework for sequence modeling by representing how hidden states evolve over time in response to inputs, offering a fundamentally recurrent alternative to attention-based architectures. Instead of computing pairwise interactions across all tokens, SSMs propagate information through a structured state update that can be computed efficiently in linear time, making them well-suited for very long sequences and streaming scenarios. The modern resurgence of SSMs began with the Structured State Space sequence model (S4) [62], which introduced stable diagonal-plus-low-rank parameterizations enabling long-range memory and efficient convolutional implementations. This foundation has since driven the development of highly expressive, scalable architectures such as Selective SSMs (Mamba) [63], RWKV [64], H3 [65], and RetNet [66], which have evolved into competitive sequence learners capable of matching or surpassing transformer performance in long-context tasks while offering significant advantages in scalability and memory efficiency. SSMs have been successfully adopted in medical imaging [67], [68].

### B. Modeling and Training

#### 1) Modeling Paradigms

We can divide the various modeling methods for FMs into generative and discriminative/ contrastive paradigms. Generative models provide a full understanding of data and produce new examples, whereas discriminative/contrastive models excel at producing generalizable representations and making decisions. Generative models offer tools for data generation, uncertainty quantification, and discovering underlying data structure, which can be invaluable in improving medical imaging qualities. Contrastive, as a type of discriminative representation learning approach, currently dominate the pretraining for image analysis tasks such as classification, segmentation, detection, and regression, underlining predictive accuracy.

#### Generative Modeling

Variational Autoencoders (VAE) [69] exemplifies an early latent-variable generative approach that marries probabilistic models with deep learning. A VAE consists of an encoder network that maps input data to a latent distribution and a decoder network that reconstructs the data from a latent sample, trained jointly by maximizing a variational lower bound on data likelihood. This framework enables learning a deep latent representation while permitting effective and efficient inference. VAEs have been pivotal as a principled method for learning unsupervised generative models of images, offering stable training and explicit probability density estimation. Furthermore, some extensions of VAEs were proposed; e.g. β-VAE [70] for disentangled factors, VQ-VAE [71] for discrete latent spaces. VAEs generate new samples rapidly but often at a relatively compromised quality in comparison with more recent models introduced below. In generative foundation models, VAEs serve as an important method for vision tokenization/compression [72].

Generative Adversarial Networks (GANs) [73] are a major step forward for generative modeling, formulated as a minimax game between a generator that synthesizes data and a discriminator that distinguishes real from fake data. Successive innovations such as DCGAN [74], Progressive GAN [75], StyleGAN [76], and BigGAN [77] improved training stability, scale, and controllability. In medical imaging, GANs have been widely adopted for cross-modality translation [78], super-resolution [79], denoising [80], inpainting [81], and simulation [82], serving as both a critical modeling paradigm in the evolution of FMs and a practical tool in medical imaging. While GANs remain attractive for their efficiency at inference, they often suffer from training difficulties and mode collapses.

Diffusion models [83] are a recent class of generative models that achieve state-of-the-art results. These models define a forward process that gradually adds noise to every image or sample in a training dataset until it becomes pure noise. Then, a learned reverse process will remove noise gradually to synthesize a new image or sample. The seminal Denoising Diffusion Probabilistic Model (DDPM) [83] demonstrated that diffusion models can produce excellent images, typically outperforming GAN results while offering advantages like stable training and distribution coverage. A main drawback of DDPM is the computational cost: generating an image requires many iterative denoising steps, making them slower than one-shot generators like VAEs and GANs. Recent research addresses this drawback with optimized samplers which allow fewer or even just one step, such as latent diffusion techniques [84] which diffuse in a lower-dimensional latent space to speed up generation and consistency models [85] which remove noise in one or few steps. Diffusion models have rapidly been adopted in medical imaging for various tasks, such as image reconstruction and enhancement [86]. These diffusion models are based on thermodynamics and have been extended in reference to electrodynamics and other mechanisms [86].

Autoregressive (AR) [87] generative models treat data synthesis as a sequential prediction problem, modeling the joint distribution of high-dimensional data as a product of conditionals. In natural language processing, this framework underlies the next-token prediction mechanism in LLMs, where each token is generated by conditioning on all previously generated tokens [44]. In computer vision, early examples include PixelRNN [88] and PixelCNN [89], which demonstrated that images can be generated pixel-by-pixel by scanning an image field and predicting the next pixel intensity conditioned on the context. However, they are notoriously slow since all output elements are produced sequentially. Recent advances demonstrate that images can be first compressed into discrete latent codes and then modeled with a Transformer as an autoregressive sequence of tokens, as illustrated by DALL·E [90] for text-to-image generation. This paradigm has proven successful in natural language processing first and more recently for image generation [91], and even in multi-modal tasks [92].

### Discriminative Modeling

Self-supervised visual representation learning (SSVRL) exploits large-scale unlabeled images to learn features. Discriminative self-supervised methods are essential for SSVRL. These methods do not attempt to model the input distribution fully; instead, they train neural networks on pretext tasks such that solving these tasks requires extracting high-level semantic features. One prominent class of methods is contrastive learning, exemplified by methods like CPC [93], SimCLR [94], MoCo [95], and PIRL [96]. Furthermore, the teacher-student learning paradigm, such as in BYOL [97], SimSiam [98], and DINO [99], took a surprising step by removing explicit negative pairs in contrastive learning. Despite the absence of contrasting against negatives, these methods avoid collapse through their asymmetric teacher-versus-student architecture. Clustering-based methods [100], [101], [102] leverage self-supervised visual representation learning, achieving excellent results. Recently, information maximization methods emerged as a promising direction for self-supervised learning due to their simplicity without contrastive negative examples nor asymmetric design [103], [104]. Self-supervised discriminative learning enables rich feature learning from unlabeled datasets, greatly reducing the need for costly annotations in medical imaging [105], [106].

Vision-Language Contrastive Learning learns joint representations from paired images and text. These approaches, such as CLIP [107] and ALIGN [108], extend contrastive learning to multimodal data: an image and its accompanying caption form a positive pair, and mismatched image–caption combinations form negatives. By leveraging extremely large datasets of image–text pairs, these models learn remarkably general and transferable visual features. After CLIP-based multimodal pretraining, the image encoder may use zero-shot training for classification. By training on noisy but abundant web data, these models encode a rich association between visual concepts and natural language, enabling important downstream applications [107]. In the medical imaging field, analogous approaches have been extensively explored, e.g. aligning radiology images with report text, to bring the benefits of multimodal pretraining to specialized domains [109].

### Generative-Discriminative Modeling

Masked autoencoders reconstruct missing or corrupted portions of the input. A prime example in NLP is BERT [110], which learns a deep bidirectional Transformer by masking out random words in a sentence and training the model to predict missing tokens. In computer vision, the Masked Autoencoder (MAE) [111] blocks a fraction of image patches and then reconstruct them, even outperforming supervised pre-training on the same architecture for downstream tasks. The MAEs are optimized via reconstruction error rather than likelihood. As such, they straddle generative and discriminative paradigms: the training objective is generative, but the resulting encoder is typically used for discriminative tasks.

Joint-Embedding Predictive Architecture (JEPA) [112] builds on the idea of a "world model", learning by predicting future or missing higher-level representations. A recent instantiation is I-JEPA [113].Another example is data2vec [114]. These approaches define an appealing middle ground that combines embedding, alignment, and prediction, learning both generative and discriminative features across modalities to support intelligent behavior. This paradigm resonates strongly with the Bayesian brain hypothesis [115] and the minimum free-energy principle [116], which similarly view intelligence as predictive modeling of latent structure in the world.

### Reinforcement Learning

Reinforcement learning (RL) has become critical in training

FMs, providing a principled mechanism to optimize model performance beyond supervised learning. While large-scale pretraining equips models with broad linguistic and world knowledge, RL enables them to incorporate explicit evaluative signals—ranging from human feedback to verifiable task-based rewards—into their policy updates, thus facilitating goal-directed refinement of model capabilities. This transition from next-token prediction to preference-aligned optimization reflects a most important trend in foundation model research, where performance reliability, controllability, generalization, and interpretability increasingly depend on an iterative feedback loop [117], [118].

RL-based methods help ensure that model outputs adhere to human-preferred behaviors, safety norms, and interaction standards. The canonical RLHF pipeline centers on Proximal Policy Optimization (PPO) [119], paired with learned reward models derived from human preference data [120]. This KL-regularized objective enables stable policy updates while preventing divergence from a reference policy. More recently, Direct Preference Optimization (DPO) [121] has emerged as a compelling alternative, reformulating the KL-constrained preference-alignment objective into a tractable supervised-learning–style loss that eliminates the need for value estimation and on-policy sampling. Complementary formulations, such as ORPO [122], further streamline preference optimization by merging likelihood training with preference modeling. In medical imaging, these methods were applied to radiology report generation [123] and radiology question answering systems [124].

For reasoning-intensive domains, models are optimized for verifiable correctness and multi-step reasoning quality. This shift has catalyzed the adoption of Group Relative Policy Optimization (GRPO) [125] and its variants. These methods compute relative advantages among multiple sampled trajectories for each prompt, thereby avoiding explicit critics and improving stability in domains where correctness signals are sparse but reliable. Enhancements such as DAPO [126] and GSPO [127] further refine group-based policy-gradient dynamics for large-scale reasoning optimization. Beyond GRPO, a growing literature explores RL with verifiable rewards [128], tree-search–augmented RL [129], self-play–driven reasoning improvement [130], and offline RL for complex reasoning trajectories [131]. Collectively, these methods position RL as a cornerstone for advancing FMs from superficially coherent reasoning to demonstrably correct, logically structured problem solving. RL-based reasoning methods have been explored for medical imaging applications, such as medical image question answering [132] and personalized lung cancer risk prediction [133].

### C. Training Workflow

A typical training workflow of FMs involves a large-scale pre-training stage for learning generalizable representations, followed by an iterative post-training process including supervised fine-tuning (SFT) and reinforcement learning for alignment and/or reasoning. Although different types of FMs may adopt varied strategies, this two-stage paradigm remains the dominant developmental framework.

**Pre-training on Broad Data**: The pre-training stage typically leverages large-scale, heterogeneous datasets to learn robust representations and capture complex data distributions. For text and vision transformers, this involves billions of text tokens or millions of images, optimized through self-supervised objectives such as masked language modeling (BERT), next-token prediction (GPT), or contrastive alignment (CLIP). In biomedical domains, "broad data'' further includes large collections of clinical text, medical images, and multimodal corpora, enabling the development of domain-specialized models such as BioLMs [134], MedSAM [135], and M3FM [136]. In many cases, domain-adaptive pretraining is performed beforehand where models are further pretrained on large in-domain corpora (e.g., BioBERT [134] trained on PubMed) to improve handling of domain-specific terminology and semantics. The pre-training of generative models such a s diffusion models [84] and GANs [137] learns to approximate the underlying data distribution closely enough to generate realistic samples or reconstruct missing or corrupted information. This generative pre-training enables models to internalize fine-grained structural and semantic patterns, while providing additional benefits unique to synthesis-based objectives. In medical imaging, such pre-trained generative models have been successfully applied to medical image reconstruction [138], super-resolution [139], denoising [140], and robust artifact correction [141]. Despite its computational complexity, pretraining is essential to extract implicit, rich, and transferable features and knowledge for downstream adaptation [136].

**Supervised Fine-Tuning**: After large-scale pre-training, models are specialized for target domains or downstream tasks through fine-tuning on smaller, high-quality labeled datasets, typically with reduced learning rates to preserve general features. A central challenge is balancing specialization with the preservation of general knowledge. Risks of overfitting and catastrophic forgetting are addressed through several effective techniques, such as layer freezing, adaptive optimization, and parameter-efficient adaptation methods (e.g., LoRA [142]).

**Reinforcement Learning for Alignment and Reasoning**: A critical aspect in the development of FMs is alignment, which ensures that outputs are not only plausible and fluent but also accurate, reliable, and consistent with domain standards and human values. A widely adopted paradigm is Reinforcement Learning from Human Feedback (RLHF), wherein human evaluators provide preference rankings of model outputs that are distilled into a reward model, subsequently optimized via reinforcement learning algorithms [119]. More recently, reinforcement learning has been renovated to improve reasoning quality. Reasoning-oriented models such as DeepSeek-R1[143] embody this shift of reinforcement learning [125] to encourage multi-step problem solving, self-consistency, and robustness in complex decision-making tasks. This shift underscores the versatility of reinforcement learning both as an alignment mechanism and as a route to strengthen reasoning performance, an essential property for high-stakes medical imaging applications. In some training pipelines, data generated after RL is fed back into the SFT dataset, creating an iterative loop that progressively improves real-world performance [143].

### D. Computational Efficiency

The rapid scaling of FMs has intensified the need for computationally efficient training and cost-effective deployment, especially for medical imaging applications where the data are of high dimensions. Efficiency is now a foundational design goal rather than a secondary consideration. Although some architectural efficiency designs such as MoE and efficient attention mechanisms have been introduced, this section reviews critical and popular techniques used to optimize training and inference phases of large-scale models.

#### 1) Efficient Optimization

Training FMs is computationally challenging. Efficient optimization techniques address this challenge by reducing memory overhead, improving parallelism, and leveraging hardware capabilities, with some important techniques reviewed below.

**Sharded Training and Memory Partitioning**: Fully Sharded Data Parallelism (FSDP) [144] and ZeRO-based [145] optimization represent two state-of-the-art strategies for memory-efficient distributed training. These methods partition model parameters, gradients, and optimizer states across devices, enabling training of models that exceed the memory limits of individual GPUs. The ZeRO family (ZeRO-1/2/3) further allows flexible control over the trade-off between memory savings and inter-device communication.

**Parameter-Efficient Fine-Tuning**: Low-Rank Adaptation (LoRA) [146] and its variants [147] reduce the computational and storage cost of fine-tuning by introducing lightweight low-rank matrices into Transformer modules. Only these small sets of parameters are updated during training, allowing adaptation to new tasks with orders-of-magnitude fewer trainable weights.

**Efficient Attention and Kernel Implementations**: FlashAttention [148] provides an optimized, memory-efficient attention implementation that minimizes redundant data movement, directly improving training speed. Similarly, kernel fusion techniques [149] combine multiple GPU operations into a single execution step, reducing kernel-launch overhead and improving hardware utilization.

**Mixed Precision and Activation Checkpointing**: Mixed-precision training [150] with FP16 or BF16 formats reduces memory footprint and increases arithmetic throughput, particularly on modern Tensor Core devices. Activation checkpointing [151] further reduces memory use by storing only a subset of intermediate activations and recomputing others during backpropagation. Together these techniques allow larger batch sizes and deeper models to be trained with the same hardware resources.

**Gradient Accumulation**: When batch sizes exceed GPU memory capacity, gradient accumulation simulates large-batch training by aggregating gradients across several forward passes [152]. This enables stable optimization behavior without requiring large accelerator clusters.

#### 2) Efficient Deployment

Once training is complete, model deployment must balance performance, latency requirements, and hardware constraints. Efficient deployment focuses on compressing models, reducing precision, and leveraging specialized inference engines to achieve high-throughput, low-cost inference.

**Model Compression via Distillation**: Knowledge distillation [153] transfers the behavior of a large "teacher" model to a smaller "student" model. The resulting student model retains much of the teacher's predictive capability while requiring substantially fewer parameters and reduced computing. Distillation is especially effective for edge devices, mobile platforms, and latency-sensitive applications.

**Quantization for Reduced Precision Inference**: Quantization converts model weights and activations from high-precision formats (e.g., FP32) to lower-precision representations such as INT8 or INT4 [154]. Modern quantization-aware and post-training quantization techniques maintain accuracy while dramatically improving inference throughput and lowering memory usage. These methods have become standard in production-scale model serving.

**Specialized Inference Runtimes**: Recent advancements in inference system design have led to optimized runtimes tailored specifically for LLMs and multimodal LLMs.

• vLLM [155] introduces PagedAttention, an efficient memory management mechanism that improves multi-request batching and maximizes GPU utilization.

• SGLang [156] extends these ideas by incorporating structured caching and partial-decoding reuse, enabling high-throughput, multi-tenant deployment scenarios.

• Ollama focuses on ease of deployment for local environments, particularly for quantized models running on consumer hardware.

These systems provide substantial improvements over traditional deep learning servers, achieving significantly higher throughput for practical applications.

## III. MEDICAL IMAGING APPLICATIONS

In this section, we survey successes and challenges of FMs in medical imaging for various modalities and tasks as shown in Fig. 3.1, and we list the major benchmarks and medical imaging data platforms for development of FMs. Medical imaging methods often face the long-tail data scenario, caused by heavily imbalanced datasets in which many common disease cases coexist with fewer rare disease cases. Consequently, the scarcity of data for training models to accurately identify these rare cases leads to performance degradation. The few-shot setting of FMs aligns perfectly with this long-tailed scenario, serving as a versatile base for a wide range of imaging modalities, anatomies, and downstream tasks.

### A. Image Reconstruction and Enhancement

Medical image reconstruction involves solving inverse problems to recover high-quality images from incomplete or corrupted data. For example, reconstructing images from undersampled k-space signals or noisy/incomplete sinograms [157], [158], [159], [160]. FMs are increasingly researched for this purpose to improve reconstruction quality under challenging conditions for various imaging modalities. Instead of training a separate network for every scanner or protocol, a single large model can be pretrained on diverse image data and then adapted to specific reconstruction tasks. This section focuses on three important types of foundations models for image reconstruction & enhancement tasks as shown in Fig. 3.2.

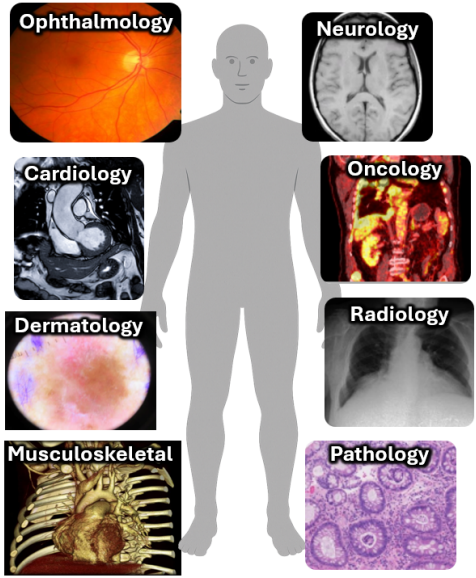

Fig. 3.1. Example applications of foundation models in different aspects of medical imaging, across different modalities, anatomies, and tasks.

### 1) Image Enhancement

These types of networks directly improve the data quality, either in a measurement or image domain, or in both domains as shown in pink in Fig. 3.2. Common examples include sinogram completion and inpainting, K-space completion, denoising [80], [161], super-resolution [79], artifacts correction, inhomogeneity correction, and harmonization networks [162]. They do not require explicitly defined forward models, since they are not solving an inverse problem.

Early attempts have demonstrated that deep learning can greatly enhance image quality in specific scenarios. However, these models typically are limited to pre-defined tasks or selected anatomic regions, and they often suffer from poor generalizability for out-of-domain tasks. Recently, there has been an increasing interest in FMs for multiple image enhancement tasks. TAMP for example leverages a physics-drive pre-training and parameter-efficient adaptation for universal CT image quality improvement in both sinogram and image domains [163].

### 2) Direct Reconstruction

Researchers began leveraging large, paired datasets to directly learn mappings from measurements (e.g., sinograms in CT, k-space in MRI) to the final clean, high-quality images in an end-to-end fashion [164]. These approaches focus on minimizing reconstruction and other losses (e.g., L1, L2, perceptual [165], adversarial [165]) during the training procedure. They are typically not iterative in nature and can use imaging system models to enforce physical constraints.

Compared to image enhancement networks, reconstruction networks can incorporate the measurement operator and observed data. For example, the "Reconstruct Anything" model proposes a universal direct inversion model by introducing a new conditioning mechanism that integrates the imaging physics through multigrid Krylov iterations [166]. This single backbone model performs multiple image reconstruction and enhancement tasks. Another prominent example is the unrolled network framework type of approach, which mimics iterative optimization algorithms (e.g., ISTA, ADMM) by embedding data consistency and learned regularization into a trainable architecture [157], [158], [159], [160]. Finally, another line of work aims to directly learn the inverse model with deep learning (i.e., not explicitly providing the imaging physics / forward model to the network), for example, iRadonMap [167], [168], hierarchical DL reconstruction [169], and AUTOMAP [170], [171]. However, making the network learn an inverse models can be challenging due to their dimensionality. We distinguish these methods from image enhancement methods since they involve a domain change between input and output (e.g., from K-space to image domain). This is still an emerging application for foundation models.

### 3) Prior Modeling

The final category of networks focuses on prior distribution modeling. Such a model can help the inversion/reconstruction process significantly. Instead of relying on paired data, they can naturally be applied to different inverse problems without fine-tuning. Most of these methods need the forward model to enforce various conditions such as tomographic data to finish tomographic reconstruction. While most of these methods are iterative, they are not the defining characteristics of the proposed method; for example, a conditional diffusion model trained to correct for different types of artifacts would still classify as an image enhancement task, despite being iterative in nature.

One popular line of work utilizes score-based models to estimate the unconditional score function of the prior distribution. Then, during inference, some form of measurement matching technique is used to condition the reverse diffusion with a closed-form, approximated measurement matching score. Some examples are DPS, Score ALD, Score-SDE, Repaint (for inpainting only), BlindDPS, DDRM, PFGMs [24], DDS [172], Blaze3DM [173], and FORCE [174]. Another line of work uses plug-and-play networks [175], [176], which use a pre-trained denoiser to regularize an iterative reconstruction process of solving inverse problems, as opposed to using a hand-crafted regularizer such as total variation [177] or wavelet sparsity [178].

While all the above-mentioned methods can potentially be used for various applications (thus qualifying as FMs), the first two categories often require retraining for new applications. In contrast, the third category offers a general framework for arbitrary inverse problems without retraining nor fine-tuning, likely at a cost of slow inference speed. Overall, while the use of FMs for tomographic reconstruction is still emerging, the trend is clear: a single large model (often generative) can flexibly handle multiple inverse problems by virtue of comprehensive prior knowledge. In major clinical areas, this means faster scans (by reconstructing from sparse data), lower radiation and contrast doses, higher image quality, and better diagnostic performance.

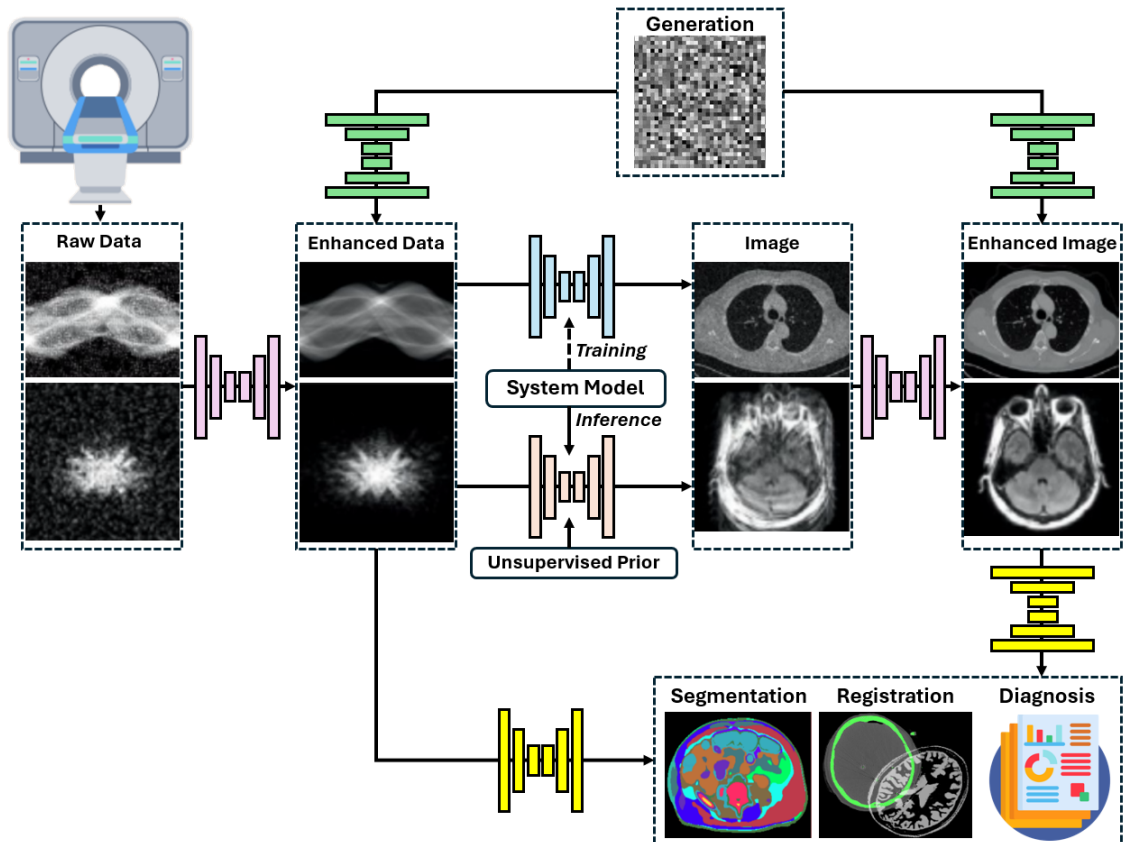

Fig. 3.2. Illustration of foundation models for different tasks in medical imaging. The center portion shows three types of foundation models for image reconstruction & enhancement. Pink: image (or data) enhancement model. Blue: direct reconstruction model. Orange: prior distribution model. The other portions show data generation models (green) and data analysis models (yellow).

### B. Image Analysis

In medical image analysis, innovative FM techniques are continually emerging in various clinical tasks. A few representative applications are summarized here.

#### 1) Classification and Regression

FMs have driven advances in medical image classification and regression, enabling diagnostic prediction and feature discovery with minimal supervision. Early deep learning models already matched expert performance in tasks like disease detection from images, but they required large, labeled datasets. FMs address this challenge by leveraging self-supervised pretraining on vast unlabeled datasets, often coupled with text, to reduce labeling efforts in various domains.

In radiology, the CheXzero model [179] was trained on hundreds of thousands of chest X-rays and their clinical reports using contrastive vision-language learning. CheXzero achieves zero-shot pathology classification, i.e., it can detect diseases that were not annotated, reaching area-under-curve (AUC) values around 0.95 for several findings on external X-ray datasets. In oncology, FMs are used to discover imaging biomarkers from radiological scans. Pai et al. trained a self-supervised encoder on 11,467 diverse tumor images [180], yielding a model that outperformed conventional supervised methods in predicting clinical biomarkers, especially in low-data regimes. In pathology, FMs are also pivotal for classification like cancer subtyping. For example, Prov-GigaPath [181] achieved state-of-the-art accuracy on 25 of 26 tasks in a pathology benchmark (covering cancer subtypes and "pathomics" predictive tasks), significantly outperforming prior methods on the majority of those tasks. Finally, regression-based approaches remain central to cancer risk assessment and survival prediction, as indicated by studies such as DeepSurv [182] and TabSurv [183].

#### 2) Segmentation and Detection

Early attempts at deep learning for segmentation has relied on task-specific models, each requiring laborious annotations [184]. FMs are redefining this area by learning generalized segmentation capabilities across diverse organs and modalities. Several studies explored how large-scale pretrained models can generalize across segmentation tasks in medical imaging. Ma et al. [185], and Noh et al [186] provide overviews of FMs for segmentation tasks, including tumor detection, organ delineation, cell segmentation, and anomaly identification.

The Segment Anything Model (SAM) [187] is a promptable

paradigm and the first FM for general-purpose image segmentation. However, applying it naively showed limited accuracy on many medical images [188]. To bridge this gap, researchers developed medical-domain variants like MedSAM [189], a FM for "universal" medical image segmentation, which was pretrained on 1.57 million image–mask pairs spanning 10 imaging modalities and over 30 disease types. It was evaluated favorably on 86 internal tasks and 60 external test tasks. Other attempts include MedLSAM [190], 3DSAM-Adapter [191], SAM-Med2D [192], SAMed-2 [193], and SAM-U [194], enhancing the sensitivity and specificity of medical image segmentation tasks.

However, promptable segmentation can be time-consuming in 3D cases for large cohort data analysis. Given that both SAM and SAM2 (video variant of SAM) handle either single or a stream or 2D images, their performance is typically worse than specially trained 3D models. To address this issue, various 3D medical segmentation models (with and without prompts) were proposed. Promptless models typically support automated segmentation of most human anatomies and major pathologies directly [195]. For example, BrainSegFounder uses SwinUNETR for 3D neuroimage segmentation tasks [196]. VISTA3D is a 3D segmentation model, whose performance was boosted by distilling state-of-the-art 2D image segmentation with supervoxels [196].

### 3) Registration

Conventional registration algorithms (e.g. deformable registration) are generic but slow [197], while recent learning-based methods are fast but tend to overfit [198], [199], [200], [201]. Emerging FMs deliver state-of-the-art results. UniGradICON[202], [203] is an early FM for medical image registration, trained on a dozen public datasets covering various anatomies. This model achieved high accuracy across multiple registration tasks (e.g., aligning brain MRI scans as well as thoracic CT scans). Similarly, Hu et al enhanced the robustness and generalizability of registration using a FM after sharpness-aware minimization [204].

Another line of work focuses on performing zero- or few-shot transfer learning using pretrained vision models. For example, DINO-REG uses the feature maps from DINO (a natural image foundation model) to compute the registration loss [205]. MultiCo3D leveraged anatomical information from the SAM model to guide registration (via aligning semantics) [206]. FoundationMorph utilizes a pretrained vision-language model to guide registration and a multi-dimensional attention module to fuse vision-language representations [205].

### C. Image Generation

Image generation especially relevant in medical imaging due to data scarcity including data imbalances (rarity of certain medical conditions and/or populations), high human-annotation cost, and patient privacy concerns [207]. By creating artificial yet realistic medical images with generative AI, one can greatly reduce the dependency on real patient data for training powerful deep learning models [208], [209], virtual clinical trials [210], [211], and training medical professionals [212]. FMs have revolutionized image generation by enabling scalable, high-fidelity synthesis across diverse modalities. Recent efforts demonstrated that generative AI can synthesize high-quality chest X-ray images, 3D MR and CT images, 2D pathology images, and so on [213], [214], [215], [216], [217], [218]. We highlight two research aspects of image generation:

Model Architecture & Conditioning: Early attempts at medical image generation typically rely on generative adversarial networks (GANs), which – while powerful - suffered from issues like mode collapse, difficulty in training, and limited sample diversity [219]. Transformers have then been used (e.g., Med-Art [220], TransMed [221], MedFormer [222]) to offer global context modeling and scalability. Recently, diffusion-type models have emerged for high-quality image synthesis [213], [214], [215], [216], [217], [218]. For example, MINIM has presented a unified text-conditioned latent diffusion model for different domains, including OCT, chest X-ray, and CT [218].

In terms of new conditioning approaches (i.e., controlled generation): one line of work uses text-to-image diffusion models adapted to medicine. For example, RoentGen [223] fine-tunes a popular latent-space vision–language diffusion model (Stable Diffusion, trained on natural images and captions) using tens of thousands of chest X-rays paired with radiology report sentences. With DiffTumor [224], CT scans can be synthesized with a liver tumor of a specified size and location. Beyond text prompts, FMs can generate images conditioned on other inputs, such as segmentation maps or existing images (image-to-image translation). Diffusion models are also used to inpaint or modify medical images in a controlled way [225].

Scalable Generation: Realistic high-resolution 3D/4D volume generation is rather challenging due to the high memory footprint required by a unified 3D framework. This is further complicated by the inhomogeneities of medical images in terms of volume dimensions and pixel sizes. GenerateCT [226] addresses this challenge by decomposing the 3D generation process into a sequential generation of individual slices. While a volume can be generated to arbitrary sizes with this approach, there are lingering concerns regarding the 3D structural inconsistencies across slices. Others attempt at direct 3D image generation: for example, MAISI achieves 512³ realistic CT image generation via latent space diffusion and Tensor splitting parallelism (TSP) [227], [228].

### D. Report Generation and Vision Question-Answering

FMs have substantially advanced automated radiology report generation. For example, FMs can generate human-readable reports from multi-modal data in structured or unstructured formats in professional and plain languages [229], [230]. Compared with earlier encoder–decoder systems, modern multimodal Transformers pretrained on datasets such as MIMIC-CXR can produce more coherent, structured, and clinically aligned reports, often with explicit sections for findings, impressions, and comparisons [231]. Recent work further applies preference-based optimization such as Direct Preference Optimization (DPO) to suppress hallucinated prior examinations and better align generated text with radiologist

TABLE 1. MAJOR DATASETS AND PLATFORMS FOR MEDICAL IMAGING.

| Name / Platform | Modality / Type | Scope (Anatomy / Task) | Annotation / Data Type | Scale |
|---|---|---|---|---|
| ChestX-ray14 [39] | X-ray | Chest diseases | Image-level labels | 100k+ |
| CheXpert [40] | X-ray | Chest (in/outpatient) | Uncertain labels | 220k+ |
| MIMIC-CXR [41] | X-ray + reports | Chest (ICU) | Full radiology reports; labels | 370k+ |
| DeepLesion [42] | CT | Whole-body lesions | Bounding boxes | 30k+ slices |
| RadImageNet [43] | CT/MRI/US | Multi-organ, multimodal | Diagnostic labels | 1.3M+ |
| BraTS [44] | MRI (3D) | Brain tumors | Segmentation masks | ~2k cases |
| EchoNet-Dynamic [45] | Ultrasound (video) | Cardiac | EF values; masks | 10k+ videos |
| TCGA [46] | WSI + CT/MRI/PET | Multicancer | Dx labels; genomics; ROIs | 20k–30k WSIs + radiology |
| PANDA [47] | Pathology (WSI) | Prostate | Gleason grades | 11k |
| ROCO[48]/MedICaT [49] | Multi-modality + text | Scientific figures | Captions; article text | 80k–200k |
| MedMNIST v2 [50] | Multi-modality | 18 organ/task datasets | Class labels | ~700k |
| MSD [51] | CT/MRI | 10 organs | Pixel masks | 2.6k volumes |
| fastMRI [52] | MRI | Knee, brain | Raw k-space; fully & undersampled | Millions of slices |
| Calgary-Campinas [53] | MRI | Brain | Fully sampled k-space | ~370 volumes |
| MRiLab synthetic MRI [54] | MRI (simulated) | Brain | Synthetic k-space | Millions |
| Mayo LDCT Challenge [55] | CT | Chest/abdomen | Full-dose/low-dose paired CT | Hundreds of volumes |
| LIDC-IDRI [56] | CT | Lung | Nodule labels; full CT volumes | 1,018 cases |
| TCIA [57] | Platform (multi-modality) | Multicancer, multi-organ | Curated datasets; Dx; segmentation; genomics | 50k+ studies |
| MIDRC [58] | Platform (X-ray, CT) | COVID-19, thoracic | Standardized imaging + metadata | 500k+ images |
| UK Biobank Imaging [59] | Platform (MRI, X-ray, fundus) | Population cohort | Structural/functional MRI; clinical data | 100k+ participants |
| NLST [60] | CT, Clinical Data | Lung cancer screening | Nodule annotations; CT series | Tens of thousands |
| OASIS [61] / HCP [62] | MRI | Neuroimaging | Structural + functional MRI | Thousands |
| AAPM Challenges [63] | Multi-modality | CXR, CT, colonoscopy, fractures | Task-specific annotations | Varies |
| PhysioNet Imaging [64] | Platform (X-ray, US, CT) | ICU and clinical cohorts | Images linked to EHR/waveforms | 100k+ |

expectations [232].

Vision question-answering (VQA) provides a novel interaction mode in which FMs answer targeted queries such as "Is there cardiomegaly?" or "What is the size of the liver lesion?" rather than producing a full report. Leveraging large-scale image–text pretraining and attention-based localization, contemporary medical VQA systems achieve strong performance across CXR, CT, MRI, ultrasound, and pathology, and can often operate in zero-shot or few-shot regimes [231]. Alignment methods like DPO have also been adapted to radiology VQA (RadQA-DPO) [233].

### E. Other Tasks

Finally, we describe how FMs may solve some applications peripherally related to medical imaging. FMs can automate patient follow-up. Using imaging data and electronic patient records, FMs can personalize messages with recommendations and appointments. FMs can identify public-health relevant disease patterns or biomarkers. For example, FMs were suggested for detecting quantitative cancer biomarkers and predicting public disease progression [234]. FMs can also be used for workflow optimization, such as image quality monitoring in a hospital, which can then alert technologists and/or physicians to image quality issues and/or incorrect protocol selection [235], [236], [237]. Other usages include automatic protocol selection and recommendation based on scout images and patient information, and automatic data management (e.g., similar case retrieval, [238]). This list seems endless, only limited by our imagination.

### F. Medical Imaging Datasets and Benchmarks

Foundation models in medical imaging rely on large, diverse, and multimodal datasets that span radiography, CT, MRI, ultrasound, nuclear medicine, digital pathology, and image–text corpora, as well as highly specialized datasets for tasks like reconstruction, segmentation, and report generation. Public platforms such as The Cancer Imaging Archive (TCIA) [239] and the Medical Imaging and Data Resource Center (MIDRC) [240] have become central hubs for standardized, curated collections across cancer imaging, COVID-19 imaging, lung screening, and multi-organ cohorts, enabling reproducible benchmarks and large-scale pretraining. In parallel, dedicated CT and MRI reconstruction datasets—such as fastMRI [241], MRiLab data [242], the Mayo Clinic Low-Dose CT Challenge datasets [243], and AAPM Grand Challenge collections [244]—provide high-quality raw k-space or projection data needed to support physics-informed foundation models. Table 1 summarizes representative datasets and platforms that underpin the development and evaluation of generalist and multimodal foundation models in medical imaging.

## IV. PERSPECTIVES

The successes of AI have been commonly attributed to the three pillars: data, models, and computing power [265]. Large, diverse, and multimodal data fuel the learning process; advanced architectures and optimization techniques extract information from data to empower generalizable models; and computing infrastructure enables the training and deployment of AI systems. Together, these three pillars have driven much of the rapid evolution of foundation models in medical imaging. However, medicine is a mission-critical domain, and medical imaging serves as the eyes of modern medicine, where errors

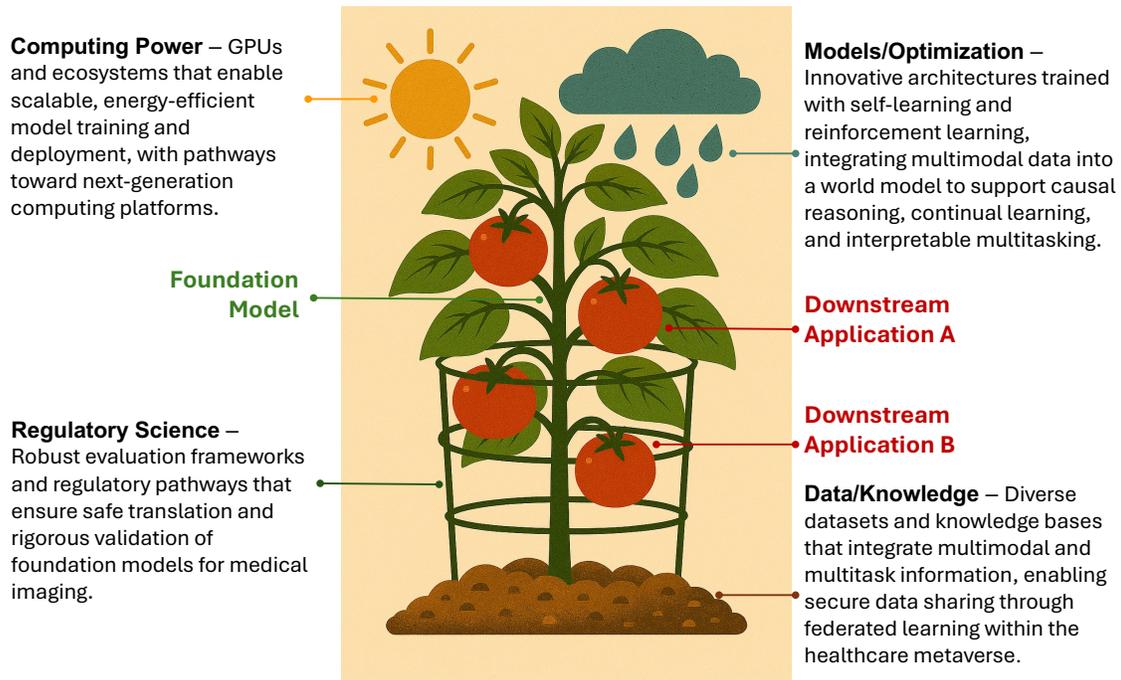

**Computing Power** – GPUs and ecosystems that enable scalable, energy-efficient model training and deployment, with pathways toward next-generation computing platforms.

**Models/Optimization** – Innovative architectures trained with self-learning and reinforcement learning, integrating multimodal data into a world model to support causal reasoning, continual learning, and interpretable multitasking.

**Foundation Model**

**Downstream Application A**

**Regulatory Science** – Robust evaluation frameworks and regulatory pathways that ensure safe translation and rigorous validation of foundation models for medical imaging.

**Downstream Application B**

**Data/Knowledge** – Diverse datasets and knowledge bases that integrate multimodal and multitask information, enabling secure data sharing through federated learning within the healthcare metaverse.

Fig. 4.1. Four pillars of foundation models for medical imaging. While three pillars for AI progress have been widely recognized, including data, models, and computing, in medical imaging the stakes demand the fourth pillar: regulatory science.

have life-altering consequences. Unlike other fields where innovation could temporarily outpace regulation, healthcare demands not only technical sophistication but also trust, safety, and accountability. This reality calls for the fourth pillar: regulatory science. Robust evaluation frameworks, fairness auditing, clinical trials, and alignment with ethics ensure that foundation models for medical imaging meet the highest standards before entering real-world clinical workflows. By explicitly adding regulatory science to the traditional triad as shown in Figure 4.1, we underline that AI in medical imaging demands more than just bigger datasets, smarter models, and faster GPUs – it requires a disciplined pathway from technical breakthroughs to clinical integration.

### A. Data/Knowledge

An unprecedented scale of datasets has been used in pretraining large AI models including foundation models. However, it is increasingly recognized that bigger is not always better, particularly in the medical domain where image subtleties, clinical representativeness, and biological complexities directly affect model outcomes [65]. While the size of a dataset is a good indicator of information content, it has become the consensus that data quality, diversity, and multimodality are equally important for development of trustworthy AI models and their clinical performance. Indeed, data quality ensures that foundation models learn meaningful patterns rather than noise or artifacts. Medical images vary widely in type, quality, style, and annotation accuracy. Low-quality or inconsistently labeled data can propagate errors through downstream tasks. Also, data diversity underpins generalizability across patient populations, scanners, and clinical settings. Many existing datasets overrepresent specific demographics or disease types. Finally, integrating imaging with clinical text, genomics, and longitudinal health records, multimodality unlocks representations that reflect the full complexity of patient care.

Despite their value, datasets remain largely fragmented across hospitals, vendors, and jurisdictions due to privacy regulations (e.g., HIPAA, GDPR), intellectual property concerns, and institutional policies. Federated learning has already served as an alternative [66]. To further improve privacy, secure computation techniques can be combined with federated learning to ensure that intermediate model updates remain encrypted [67]. Moreover, synthetic data synthesis using generative AI can fill gaps where real data are scarce, sensitive, or inaccessible. When integrated with privacy-preserving pipelines, synthetic data can be shared without compromising patient confidentiality [68].

Additionally, we envision a legally mandated framework under which medical datasets would be preserved securely during patients' lifetimes and be declassified a few decades later for research use. Such a mechanism, analogous to historical archives in other domains, would balance privacy and scientific value. This will allow future researchers to inherit a comprehensive, ethically sourced repository for studying disease evolution, health trends, and long-range biomedical questions. This means that data governance must not only protect individuals now but also enable transformative discoveries for generations to come.

Beyond sheer data volume, incorporating medical knowledge is emerging as a critical enabler for medical foundation models [69]. Knowledge graphs capturing biomedical ontologies, disease–symptom–treatment relationships, and imaging–genomics linkages can guide representation learning toward clinically meaningful concepts. Likewise, retrieval-augmented generation (RAG) allows models to query curated clinical databases, imaging atlases, and knowledge bases during inference, providing verifiable yet up-to-date context for predictions [70].

### B. Models/Optimization

Currently, autoregressive, and diffusion-type models represent two major paradigms. Large language models (LLMs) work via autoregression and are very successful. On the other hand, diffusion-type models have demonstrated exceptional performance in image, video, and multimodal generation. These paradigms can be in contrast: tokenization versus transformation, next-token prediction versus field-based generation, symbolic reasoning versus perceptual changing, and semantic analysis versus manifold learning. Autoregressive models may suffer from exposure bias/error accumulation, but diffusion-type models face computational costs due to iterative sampling, though new solvers and latent-space parameterizations mitigate this issue. Now, a growing trend is to unify both paradigms [71], [72].

Up to today the Transformer architecture remains the mainstream, but breakthroughs are needed for performance boost. Emerging architectures such as Mamba models show the ability to capture long-range context with sub-quadratic complexity [73]. Further improvements like MambaExtend enhance its long-context capabilities via a training-free scaling calibration, enabling up to 32 times longer context windows with minimal computational overhead [74]. Also, differentiable reasoning engines and hybrid neuro-symbolic architecture are emerging to integrate symbolic knowledge with deep learning [75]. Efforts are further made toward brain-inspired architectures that work with novel links, loops, and emerging behaviors [76]. Inspiration from the human brain suggests future artificial neural network architectures with modularity, recurrent connectivity, memory, predictive coding, and cross-modal integration, perception, and reasoning in fast and slow modes.

While data-driven techniques offer immense value, they cannot fully replace the role of physics-based models. When the underlying physics is well understood and models accurately approximate reality in a generalizable way, physics-based approaches will consistently outperform purely data-driven ones. Therefore, hybrid architectures that combine physics modeling with deep learning—such as physics-informed foundation models (FMs)—are likely to deliver the most robust and powerful solutions. When grounded in first principles, these FMs become even more foundational.

Since a neural architecture is nothing but a computational prototype, we must train it to optimize its parameters and performance, including pre-training, training, and post-training (during test-time/inference) [77]. This is critical but highly nontrivial, as it demands non-convex optimization. Techniques such as reinforcement learning with human or tool feedback and fine-turning in various forms can guide models toward desirable outcomes and meaningful decision-making [78]. Dynamic optimization strategies, including test-time training and adaptation, will be essential to maintain performance under distribution shifts [79].

As medical imaging foundation models evolve, a natural tension lies between generalist models and specialist models. Generalist models benefit from large-scale, multi-modal pretraining, offering transferable representations and a unified inference pipeline across diverse tasks [80]. On the other hand, specialist models, often built through parameter-efficient fine-tuning of a generalist backbone, can achieve higher accuracy, regulatory clarity, and workflow integration for targeted use cases such as lung nodule tracking or cardiac function assessment [81], [82]. Future ecosystems will likely adopt a combined paradigm, where a robust generalist foundation provides shared representations, while specialist derivatives deliver precision, interpretability, and regulatory compliance for mission-critical clinical endpoints. We believe that interactions between specialist and generalist models exemplify bottom-up and top-down methods, defining the dynamics of medical AI.

### C. Computing Power

The rapid advancement of medical imaging foundation models depends critically on high-performance computing resources, with GPU being the main work horse. The pace of computational innovation, from NVIDIA's cutting-edge architectures to emerging specialized AI accelerators, has enabled ever-larger models, shorter training cycles, and faster inferences, but it also demands investment to stay competitive. Initiatives such as New York's EmpireAI exemplify forward-looking efforts to enable and democratize access to powerful computing resources, fostering partnerships across the state. As the first state-initiated AI-oriented computational infrastructure, Empire AI [83] brings together nine universities in New York to operate a shared high-performance computing resources. Designed to become a unique academic computing platform, Empire AI has demonstrated strong results with its first-generation Alpha version. The upcoming Beta version represents a major boost, with 7X speedup and 20X acceleration in inference, capable of training multi-trillion-parameter models, Beyond the Beta version, the Buffalo supercomputing facility is expected in 2027 to deliver orders of magnitude greater computational power. Meanwhile, tech giants have active projects like Stargate, highlighting the global best-in-class resources. For medical imaging, forging synergies between universities, hospitals, industry stakeholders, and federate initiatives like EmpireAI will be key to ensuring that cutting-edge computational capacity translates into real-world clinical innovations.

In addition to GPUs, emerging computing paradigms hold promises for development of medical imaging foundation models. Quantum computing has the potential to revolutionize foundation models by accelerating large-scale training, enabling quantum-inspired architectures, and unlocking generative and/or discriminative capabilities for complex, high-dimensional problems. As quantum hardware matures, hybrid quantum-classical FMs could become essential, paving the way for breakthroughs in medical imaging and beyond. Neuromorphic computing, inspired by spiking neural networks and event-driven architectures, offers ultra-low-power inference and real-time edge intelligence. Optical computing harnesses photonic interconnects and analog light-based operations to achieve massively parallel, energy-efficient matrix computations beyond the scaling limits of electronic chips. At another frontier, synthetic biological intelligence does computation using engineered cellular or molecular systems, opening a radically new substrate for learning and decision-making under biochemical constraints. As these technologies mature, hybrid computing ecosystems that integrate electronic, photonic, neuromorphic, and biological processors could deliver orders-of-magnitude improvements in speed, energy efficiency, and adaptability.

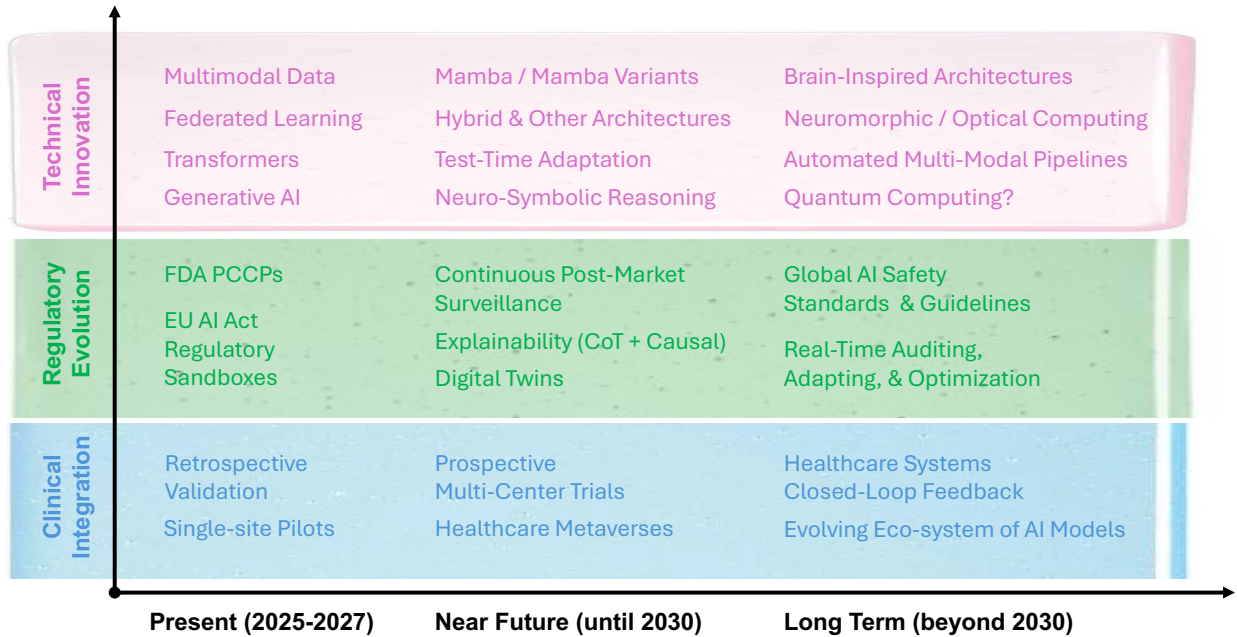

Fig. 4.2. High-level roadmap for development of medical imaging foundation models underpinned by the four pillars, highlighting how technical, regulatory, and clinical advances converge to enable trustworthy, high-performance AI systems, while ensuring their safe, ethical, and viable deployment for healthcare benefits.

### D. Regulatory Science

The rapid development of medical imaging foundation models (FMs) has outpaced existing regulatory frameworks, underscoring the urgent need to develop new rules tailored to address their unique challenges. While the U.S. FDA has begun adapting its existing AI/ML-based software regulations, such as predetermined change control plans and Good Machine Learning Practice (GMLP), these general frameworks do not fully account for the unique features of medical foundation models. Unlike traditional task-specific AI systems, foundation models are pre-trained on vast, heterogeneous datasets and can be fine-tuned or prompted for a range of downstream tasks. This introduces new regulatory challenges. Given the mission-critical nature of medical applications, it is essential to develop a dedicated regulatory science strategy for foundation models to address generalizability, explainability, monitoring, and so on.

To embed explainability in foundation models, a promising approach is to synergistically combine Chain-of-Thought (CoT) reasoning [84] and causal analysis [85]. CoT reasoning offers a narrative, step-by-step explanation of how a model arrives at a decision, helping users interpret its internal logic. However, these reasoning traces can be post hoc rationalizations rather than genuine causal mechanisms. On the other hand, causal analysis aims to uncover the underlying cause-and-effect relationships driving a model's predictions. While more rigorous, causal models alone may lack the intuitive transparency needed for clinicians to trust and adopt the system. We advocate that integrating these two complementary approaches by aligning the CoT explanation with the model's learned or inferred causal structure [86]. In principle, this coupling can produce a dual-layered explainability framework that is both human-interpretable and epistemically sound. Such an approach not only enhances the trustworthiness of medical AI systems but also provides regulators with a principled method to evaluate explainability claims. In the context of foundation models where tasks and data may shift significantly over time, this approach offers a scalable and rigorous way to monitor model reasoning across clinical scenarios.

A cornerstone of regulatory oversight is ensuring generalizability under distribution shifts. Prospective multi-institutional benchmarks should stress-test models across raw acquisition, reconstruction, enhancement, and diagnostic tasks to quantify variability introduced by hardware differences, imaging physics, and clinical protocols [87] [88]. Techniques such as domain adaptation, test-time training, and federated or cross-site training reduce sensitivity to site- or device-specific artifacts. Subgroup performance disparities can undermine clinical trust, and can be measured with disaggregated metrics, dataset design through balanced cohort selection, and augmentation for underrepresented groups. Equally important is that AI models should produce well-calibrated confidence estimates [89] and abstain from predictions when uncertainty is high, providing an indicator for potential deterioration of model performance over time. Also, retrieval-augmented generation and tool grounding help ensure that diagnostic claims remain linked to verifiable evidence rather than hallucinated content [90].

Regulatory science treats model deployment as the beginning of oversight. Clinical integration requires continuous monitoring for data drift, adversarial inputs, and performance regressions, coupled with shadow mode evaluations before live deployment. The FDA's Predetermined Change Control Plans (PCCPs), the EU AI Act, the UK's MHRA SaMD framework, and the NIST AI Risk Management Framework converge on principles for pre-authorized updates, locked evaluation datasets, rollback procedures, and post-market surveillance. In this monitoring process, transparent governance underpins all regulatory stages. Model cards, data sheets, and stratified performance specifications by site, modality, and patient subgroup create an auditable evidence base. Documenting training datasets and steps, algorithmic changes, calibration methods, and real-world performance metrics supports reproducibility and facilitates global regulatory harmonization.

### E. Concluding Remarks

The convergence of data/knowledge, models/optimization, computing power, and regulatory science is redefining medical imaging foundation models. Further success demands high-quality, multimodal data, innovative architectures, sustainable computing ecosystems, and contemporary regulatory frameworks. Future foundation models will likely combine generalist representations across modalities and tasks with specialist derivatives fine-tuned for high-stakes clinical endpoints, supported by advances in federated learning, privacy-preserving synthetic augmentation, and retrieval-augmented reasoning to break down data silos. At the same time, brain-inspired architectures, physics-informed generative models, and emerging computing paradigms promise major gains in efficiency and capability. Yet, clinical translation will ultimately hinge on rigorous and transparent governance and continuous post-deployment oversight. By uniting technical breakthroughs with ethical and regulatory rigor, medical imaging foundation models can evolve into future healthcare systems, being powerful, trustworthy, and impactful.